# The Tara Polaris scientific vision: Advancing our understanding of the central Arctic Ocean to better address life in the Earth System

M. Babin*[1, 19], L. Karp-Boss[2], C. Bowler[3, 19], M. Ardyna[1, 19], J. M. Flores[4], M. Geoffroy[5,18], J.F. Ghiglione[6, 19], K.S. Law[7], M. Nicolaus[8], B. Rabe[8], J. Schmale[9], S. G. Acinas[10], J. W. Deming[11], P.E. Galand[12, 19], T. Linkowski[13], C. Moulin[13, 19], É. Pelletier[14, 19], I. Polyakov[15], J.-C. Raut[7], S. Rysgaard[16], M. Vancoppenolle[17], and R. Troublé[13, 19]

1. Takuvik International Research Laboratory, CNRS (France) — Université Laval (Canada), Département de biologie and Québec-Océan, Université Laval, Québec, QC, Canada

2. School of Marine Sciences, University of Maine, Orono, 04469, Maine, USA

3. Institut de Biologie de l'École Normale Supérieure, Ecole Normale Supérieure, CNRS, INSERM, Université Paris Sciences et Lettres, 75005 Paris, France

4. Weizmann Institute of Science, Department of Earth and Planetary Science, Rehovot, Israel

5. Centre for Fisheries Ecosystem Research, Fisheries and Marine Institute, Memorial University, St. John's, NL, Canada

6. Laboratoire d'océanographie microbienne (UMR7621), Sorbonne Université, CNRS, Observatoire Océanologique de Banyuls, Banyuls sur Mer, France

7. Laboratoire Atmosphères, Observations Spatiales (LATMOS)/IPSL, Sorbonne Université, UVSQ, CNRS, Paris, France

8. Alfred-Wegener-Institut Helmholtz-Zentrum für Polar- und Meeresforschung, Bremerhaven, Germany

9. Extreme Environments Research Laboratory, École Polytechnique Fédérale de Lausanne, EPFL Valais Wallis, Sion, Switzerland

10. Department of Marine Biology and Oceanography, Institute of Marine Science (ICM-CSIC), Barcelona, Spain

11. School of Oceanography and Astrobiology Program, University of Washington, Seattle, WA, USA

12. Laboratoire d'Ecogéochimie des Environnements Benthiques, Sorbonne Université, CNRS, Observatoire Océanologique de Banyuls, Banyuls sur Mer, France

13. Tara Ocean Foundation, Paris, France

14. Metabolic Genomics, Genoscope, Institut de Biologie François-Jacob, CEA - CNRS - Université Paris Saclay, Evry, France

15  University of Alaska Fairbanks, Fairbanks, USA

16  Department of Biology, CIFAR, Aarhus University, Aarhus, Denmark

17  Laboratoire d'Océanographie et du Climat, CNRS/IRD/MNHN, Sorbonne Université, Paris, France

18  Department of Arctic and Marine Biology, UiT The Arctic University of Norway, 9019 Tromsø, Norway

19  Research Federation for the Study of Global Ocean Systems Ecology & Evolution, FRA2022/Tara GOSEE, 3 rue Michel-Ange, 75016 Paris, France

\*  **Corresponding author**

**Abstract**

The Arctic Ocean is currently experiencing, at the forefront of global concerns, the pressures of climate change and global pollution. To boost our ability to understand the state of this ecosystem, its evolution in this context and its resilience, the Tara Ocean Foundation has built the *Tara Polar Station* (TPS), intended to become a permanent observatory of the central Arctic Ocean. The objective of this initiative is threefold: to deepen our knowledge of the foundations of life in an ice-covered polar ocean, to better understand the dynamics of the coupled ocean-ice-atmosphere system and the role of living organisms, and to identify long-term trends in the main characteristics of the Central Arctic Ocean ecosystem resulting from global change. In this article, we describe the vision that guided the development of the Tara Polaris scientific programme, and more specifically the first of ten transpolar drifts that will be undertaken over the next 20 years aboard the TPS (Tara Polaris I, II, III, etc.). The research activities of the Tara Polaris I expedition will be grouped under four specific but interrelated themes: biosphere-atmosphere interactions, epi- and mesopelagic life in an ice-covered ocean, life in sea ice, and pollution. In addition, a theme that cuts across all environmental compartments and disciplines, and is implemented on all Tara Polaris expeditions, is the establishment of an observatory that will monitor the main sentinels of this ecosystem. This umbrella article introduces these different themes, which are then described in more detail in four other articles in this Special Feature, in addition to an article describing the technical characteristics of the TPS.

**The Arctic is at risk under pressure from global change**

The Arctic Ocean is currently experiencing some of the most dramatic changes associated with global change:

- Air temperatures have risen 3–4 times faster than the global average (Ballinger et al., 2020; Thoman et al., 2020);

- The Arctic Ocean has lost 40% of its summer sea ice, and the remaining ice pack is changing in physical and ecological properties (Sumata et al., 2023);

- The resulting reduction in albedo and profound changes in the nature of ocean-atmosphere exchanges are impacting atmospheric and oceanic circulation, which may have consequences for regional weather and climate, including lower latitudes (Francis and Vavrus, 2021);

- The acceleration of the water cycle has led to an increase in river flow and precipitation over the Arctic Ocean, increasing the supply of freshwater (Haine et al., 2015), in turn affecting the ventilation of the Arctic Ocean (Gerke et al., 2024);

- Major changes in the cycling of elements, such as carbon and nitrogen, which are closely associated with the functioning of marine ecosystems (Li et al., 2009; Lannuzel et al., 2020);

- Increasing transport of atmospheric particles into the Arctic Ocean from boreal fires, mineral dust and anthropogenic emissions over continental regions (Meinander et al., 2025);

- Accumulation of pollutants such as mercury, persistent organic pollutants and nano- and microplastics, carried by the atmosphere, rivers and ocean currents (Bergmann et al.; Morris et al., 2022); and

- Increasing human activity in the Arctic Ocean, particularly due to shipping, tourism and resource extraction (Huntington et al., 2023).

The most dramatic of these major disturbances is a profound transformation of the sea ice cover - an overall thinning, a new open-water season in the seasonal ice zone, and a rapid stepwise disappearance of "old" multi-year ice. Because sea ice acts as a nexus at the air-sea interface in the polar seas, its transformation fundamentally jeopardizes the current functioning of the Arctic's coupled atmosphere-ice-ocean system, with climatic consequences at the local, regional and planetary scales. Sea ice is also at the heart of the functioning of the highly unique Arctic marine ecosystem. It internally harbours a rich community of microorganisms, and it is an integral component of the underlying pelagic ecosystem (Babin et al., 2025). Cumulatively, ongoing changes have the potential to profoundly alter the functioning of the Arctic marine ecosystem and its community composition at all trophic levels (Post et al., 2009; Ardyna and Arrigo, 2020). For example, we are already seeing an increase in pelagic species of zooplankton and fish to the detriment of demersal fish and molluscs in the Bering Sea (Grebmeier, 2012) and the Barents Sea (Fossheim et al., 2015). Boreal species of microalgae are increasing in abundance (Neukermans et al., 2018), including some that are toxic (Anderson et al., 2021). According to Inuit communities, Atlantic salmon are increasingly present in certain areas of the Canadian Arctic (Bilous and Dunmall, 2020). The loss of multi-year sea ice, and the profound changes in seasonality associated with these changes in pack ice, pose an existential threat to sea ice-dependent organisms. The dramatic changes underway in the Arctic and the impacts on the marine ecosystem, which are being added to an already long list at a breathtaking pace, are fuelling growing environmental insecurity.

Ice over the ocean and very cold air above are two features that were found throughout the world's oceans during the long ice ages of the 'snowball Earth' that occurred hundreds of millions of years ago. Similar features are also found today on other planets harboring a liquid ocean covered by ice, such as the moons of Jupiter and Saturn, Europa and Enceladus, respectively. The life forms, metabolic pathways and evolutionary mechanisms observed today in the sea ice and water column of the Arctic Ocean may therefore shed light on the role that sea ice might have played in the evolution of life on Earth, when life was still limited on the continents, and on the possible presence of living organisms on other oceanic worlds. The loss of multiyear sea ice will compromise our ability to study and understand past glaciation and the potential existence of life on other ocean worlds, underscoring the pressing need to explore and understand this unique ecosystem.

**The central Arctic Ocean: A highly variable seasonally driven system**

In the immediate region of the North Pole, the central Arctic Ocean (CAO) basin is around 4,000 m deep, reaching almost 5,500 m in parts (compared to the North Atlantic's 8,400 m). The CAO is a convergence zone for continental and marine water masses, as well as for air masses. Although distant, it is connected to lower latitudes and under the pressure of the heavily anthropized northern hemisphere. Its state therefore informs us about how humanity is altering the Earth.

Once a place of fascination for explorers, the CAO basin has been the focus of attention for polar and climate scientists for the past few decades, despite the immense logistical challenge of studying it. While there are many permanent communities and land-based research stations surrounding the Arctic Ocean, enabling a large number of observations, including community-based, to be collected on the surrounding geo- and ecosystems, the CAO remains difficult to access, particularly during the winter. The use of research icebreakers has made it possible to carry out a number of expeditions at the northernmost latitudes of our planet, but most of these are seasonally limited to summer. To make up for the lack of observations during other seasons, which are at least as important for the functioning of CAO, researchers have used temporary scientific stations deployed on pack ice or drifting pieces of glaciers. The long-running Soviet/Russian North

Pole drift station series of the 20th century, which began in 1937 (Shirshov and Fedorov, 1938), did much to highlight the central role of the coupled ocean-ice-atmosphere system in the Arctic, particularly in terms of its physical functioning. Ships, locked in and drifting with the pack ice, as Nansen had done at the end of the 19th century (Nansen, 1902), have also been used as platforms for year-round observations. This approach was first achieved in the modern era by the research schooner *Tara*, which drifted with the pack ice in the CAO from 2006 to 2008, as part of the EU project DAMOCLES (Developing Arctic Modeling and Observing Capabilities for Long-term Environmental Studies) scientific program led by Jean-Claude Gascard. Given the small size of the schooner and the small number of participants it carried (8–11), the science program was limited in scope and focused on frazil ice formation, radiation budgets, standard atmospheric measurements and marine vertical profiles of salinity and temperature (Nicolaus et al., 2010; Gascard et al., 2015). In 2019, the international MOSAiC project (Multidisciplinary Drifting Observatory for the Study of Arctic Climate) launched the most ambitious drifting expedition in the CAO aboard the RV *Polarstern*. The operations carried out during this expedition enabled the physical interactions between the various components of the atmosphere (radiative energy, dynamics, aerosols and clouds) (Shupe et al., 2022), the pack ice (snow, thermodynamics, deformations) (Nicolaus et al., 2022) and the ocean (shear at the ice-ocean interface, turbulent mixing, vertical structure, advection, sub-/mesoscale eddies) (Rabe et al., 2022) to be analysed over the four seasons, on spatial scales ranging from metres to hundreds of kilometres. The MOSAiC research programme included, in contrast to most previous year-long CAO expeditions, biogeochemical and biological components designed to better understand certain aspects of the seasonal nature of this ecosystem, particularly in winter (Fong et al., 2024). In 2022, Russia resumed their North Pole series of transpolar drifts, now using a new dedicated ship, the *Severny Polyus*, instead of containers.

Organisms in the CAO have presumably adapted to the extreme seasonal variations in physical characteristics of their environment. Solar radiation, which is essential for the growth of photosynthetic primary producers, is very low or virtually non-existent during the polar night, which lasts almost 6 months at the North Pole, and then peaks at a planetary maximum in photons per day on June 21 in the middle of the more than 6-month midnight sun period. We now know that even in the depths of the polar night, the CAO ecosystem remains very active (Berge et al., 2015). This period could be decisive in the life cycle of a number of organisms, even though we know only very little about how they live and survive during the polar night. The lengthening or shortening of daylight around the spring and autumn equinoxes is one of the most spectacular biotope transitions, going from 0 to 24 hours between March 4 and March 31 at 85°N. While the ocean offers a relatively stable environment in terms of temperature, such stability is not the case within sea ice, which can subject the microorganisms living within to temperatures ranging from approximately −2°C to below −20°C. These temperature variations, and more broadly the dynamics of sea ice, lead to significant changes in its microphysical structure that sympagic (in ice) microorganisms are confronted with directly. It is during the fleeting autumn freeze-up period that changes in the microphysical structure of sea ice are at their most extreme, when the ice macroscopically transforms from a granular, quasi-fluid state (frazil) to a highly structured, consolidated form. During this physically tumultuous period, the microorganisms present in seawater appear to colonize the ice en masse (Gradinger and Ikävalko, 1998), following mechanisms that are still poorly understood. In fact, many questions remain completely unanswered concerning the seasonality of the CAO ecosystem, such as the role of mixotrophy in the survival of photosynthetic organisms during the polar night, the role of chronobiology in the temporal sequence of events, and the existence of metabolic pathways specific to sympagic and pelagic lifestyles.

Seasonality is a fundamental dimension of the functioning of the coupled ocean-ice-atmosphere system and of the Arctic marine ecosystem. Past annual expeditions in the CAO have not fully addressed this dimension, particularly for the ecosystem, if only because the freeze-up has never been sampled intensively. Moreover, the intimate connection between this coupled system and the ecosystem of the CAO has never received dedicated attention. Living organisms are present in all three compartments of the coupled system. Many move from one to another. They are subject to a variety of physico-chemical conditions, but they also impact them in return.

Although the few previous full-year multidisciplinary expeditions, in the CAO or elsewhere in the Arctic Ocean (e.g., Canadian Arctic Shelf Exchange Study (CASES), Surface Heat Budget of Arctic Ocean (SHEBA), Circumpolar Flaw Lead study (CFL), *Tara* Arctic and MOSAiC), made major contributions to advancing our knowledge of the Arctic Ocean (Perovich and Moritz, 2002; Fortier and Cochran, 2008; Barber et al., 2010; Gascard et al., 2015; Shupe et al., 2022), these initiatives were one-shot efforts. Likewise, seasonally limited drifting expeditions, such as the Norwegian young sea ICE project (N-ICE2015) (Granskog et al., 2018; Graham et al., 2019), the Arctic Ice Dynamics Joint Experiment (AIDJEX) (Untersteiner et al., 2009), the Arctic Internal Wave Experiment (AIWEX) (Levine et al., 1985), and the Arctic Summer Cloud Ocean Study (ASCOS) (Tjernström et al., 2014), carried out topically focused or multidisciplinary observations from ships and adjacent ice floes. Interannual variability is such that drawing broader conclusions from a single expedition is not completely robust. Long-term mooring-based programs in the central Arctic (the Nansen and Amundsen Basins Observational System, NABOS) (Pnyushkov and Polyakov, 2022), and the Beaufort Gyre Exploration Project (BGEP) (Proshutinsky et al., 2019) have been essential to documenting physical impacts of climate change over time, especially the Atlantification of the Arctic, but have been more limited in their ability to identify linked ecosystem effects. To date, no effort has been able to document in situ, multi-annual trends in both the CAO coupled ocean-ice-atmosphere system and the Arctic marine ecosystem that are directly associated with global change, distinct from natural variability. Although remote sensing provides highly valuable information about clouds, aerosols, and the state of the Earth's surface (e.g., snow, sea ice, and the ocean surface) essentially below 80°N, it has limitations in terms of resolving critical vertical features and penetrating the ocean. Establishing a long-term observatory and research station in the CAO is also particularly relevant in light of the 16-year international agreement, signed in 2021, to ban commercial fishing in the CAO until a better understanding of the properties and dynamics of this complex ecosystem has been gained. Growing industrial interests, such as shipping or resource extraction, further point to the critical need to increase observing capacities and understanding of the complex environmental processes and functionality of the CAO, to ensure responsible policy decision-making.

**The *Tara Polar Station* science program concept**

In 2014, the Tara Ocean Foundation initiated the development of the *Tara Polar Station* (TPS), built between 2023 and 2025 and ready for its first transpolar drift in 2026. Inspired by the first transpolar drift experiment in the CAO from September 2006 to January 2008 aboard the schooner *Tara* (Gascard et al., 2008), as well as the productive scientific results (see below) of the long global *Tara* Oceans expedition (2009–2013) (Sunagawa et al., 2020), the primary intention behind the creation of the TPS was to ensure a long-term scientific presence in the CAO under safe, environmentally friendly (minimal footprint) and cost-effective conditions. The TPS was conceived to serve both as an observatory for the CAO and an international polar station, similar to the International Space Station. Starting in January 2018, a group of scientists formed around the TPS initiative began to develop the outlines of a scientific programme tailored to the station's deployment strategy and capabilities, but also unique in the landscape of current and planned research activities in the CAO. Made up of a combination of Arctic experts and representatives of the *Tara* Oceans consortium, this ever-growing group has produced the Tara Polaris scientific programme summarised below and presented in detail in the accompanying set of articles in this Elementa special feature.

Technical information about the TPS and its construction are provided in Moulin et al. (2025). In short, the TPS is a modestly sized, oval-shaped vessel, measuring 26 m in length and 16 m in beam. Like the schooner *Tara*, she was designed to withstand being trapped in the ice of the CAO. She has 4 decks, including a main deck with a surface area of around 300 m$^2$, half of which is dedicated to laboratory space. A moon pool in the center of the platform offers direct access to the ocean. The laboratories are designed to carry out a range of oceanographic and cryospheric work, as well as biological sampling, measurements and experiments. The TPS also offers the possibility of deploying an atmospheric measurement programme. The station will be occupied by a crew of 6 members (sailors, a physician and a media person), and a team of 6 or 12 scientists, depending on the season. The vessel will be deployed on a recurring basis in the CAO to carry out at least 10 transpolar drifts over the next 20 years. Each drift (termed Tara Polaris I, Tara Polaris II, and so on) will begin during freeze-up in the eastern section of the CAO and continue until ice break-up, and possibly through the subsequent freeze-up in Fram Strait, while being carried along with the pack ice by the transpolar drift. The programme meets the need for long-term monitoring of the CAO, and its framework is designed to ensure continuity in the measurements essential for detecting multi-annual trends in critical sentinels of global change. At the same time, our approach allows the program to evolve over successive expeditions, as new research questions and needs emerge. This scientific program has been built through a close association between biologists from the *Tara* Oceans consortium and Arctic specialists from multiple disciplines, covering the ocean, the sea ice and the atmosphere.

**The Tara consortium capital**

Since 2003, the Tara Ocean Foundation has enabled several research expeditions, the most successful to date, in terms of contributions to fundamental understanding of planktonic ecosystems in the ocean, being the *Tara* Oceans mission. From 2009 to 2013, this programme performed a global scale expedition across the world's ocean, including a circumnavigation of the Arctic Ocean and sampling in the Southern Ocean. Overall, the team of scientists in the *Tara* Oceans expedition systematically collected eco-morpho-genetic data at 210 sites and at depths ranging from 5 m to 1000 m, covering most of the biogeographic provinces of the world's oceans. In total, >35,000 seawater and plankton samples were collected and deposited at different partner laboratories (Karsenti et al., 2011; Pesant et al., 2015). By the time of completing the collection of samples, *Tara* Oceans constituted 19 partner institutions that began generating, organizing, and analysing the vast amounts of data derived from these samples.

Foundational resources from the project were described in a special issue of Science (Brum et al., 2015; de Vargas et al., 2015; Lima-Mendez et al., 2015; Sunagawa et al., 2020). Additional resources now include an atlas of more than a hundred million genes from microbial eukaryotes (Carradec et al., 2018) and an expanding collection of single cell genomes from uncultured yet abundant ocean microbes (Seeleuthner et al., 2018; Sieracki et al., 2019). Other publications from the consortium have emphasized the *Tara* Oceans eco-systems biology approach that aims to leverage the mechanistic cellular approaches of systems biology and the ecological community approaches of ecosystems research. Highlights include the realization that radiolarians are present in numbers at least as high as zooplankton (Biard et al., 2016), that viruses may drive rather than short-circuit the global ocean carbon pump (Guidi et al., 2016), that most eukaryotic diversity resides in symbionts and parasites (de Vargas et al., 2015), that biotic interactions are much more important than heretofore realized (Lima-Mendez et al., 2015), and that marine plankton communities display latitudinal gradients (Ibarbalz et al., 2019). The following summer, the *Tara* Oceans Polar Circle expedition contributed a comprehensive pan-Arctic dataset of uncultured microbial genomes with the identification of the potential prokaryotic arctic sentinels (Royo-Llonch et al., 2021).

Beyond the consortium, the public availability of the datasets has ensured additional fundamental insights in ecology, evolution, and Earth System functioning. *Tara* Oceans most recent ground-breaking contributions include estimates of the distribution of the world's biomass (Bar-On et al., 2018); discovery of new origins of mitochondria (Martijn et al., 2018); identification of heterotrophic nitrogen-fixing bacteria in the ocean (Delmont et al., 2018); discovery of a new class of microbial rhodopsins with potential application to optogenetics (Pushkarev et al., 2018); and discovery of 'huge' marine phages with new CRISPR/Cas systems (Al-Shayeb et al., 2020; Pausch et al., 2020).

*Tara* Oceans has also contributed to establishing standards for sample collection, processing and analysis (ten Hoopen et al., 2015; Gorsky et al., 2019), and to outreach and societal impact through numerous initiatives with school children, artists and policy makers (Sunagawa et al., 2020).

**Five themes for the first TPS expedition: Tara Polaris I**

The scientific programme of the TPS aims to improve understanding about the functioning of the Arctic biogeosphere within the coupled ocean-sea ice-atmosphere system. Although Arctic-focused, Tara Polaris has been designed to pursue two ideals: 1) to understand the origins, diversity and functioning of life on Earth, and 2) to identify the driving forces behind global change (climate change and other changes) and its impact on the living world. This intention is why the programme has a strong discovery dimension, as well as a dimension of understanding the dynamics of the Earth system and its physical and biological components, in the context of global change. Certain specific but highly diversified research themes are thus set to change from one expedition to the next, while the activities designed to identify long-term trends through the 20+ year deployment strategy will be adapted to this perspective and will cut across all the research themes.

After providing some background on specifically how we will address the coupled ocean-sea ice-atmosphere system and describing the scales covered, we briefly present four scientific themes specific to Tara Polaris I and then the long-term observational theme, which is transversal to all others. The five of them are presented in detail in Geoffroy et al. (2025), Ghiglione et al. (2025), Nicolaus et al. (2025), Schmale et al. (2025) and Vancoppenolle et al. (2025).

*The coupled ocean-sea ice-atmosphere system in a nutshell*

Decades of research have demonstrated the fundamental role of the coupling between the ocean and the atmosphere in the functioning of the Earth System. In frozen seas, the emblematic presence of sea ice substantially alters this coupling. Although it represents only a relatively thin film at the interface between the ocean and the atmosphere (from a few tens of centimetres to a few meters of ice), compared to thousands of metres of air above and thousands of metres below in the water column of the CAO, sea ice is the central element of what can be called the coupled ocean-ice-atmosphere system. The functioning of the Arctic Ocean can only be understood in terms of this coupled system.

Sea ice is in tight thermodynamic interaction with the atmosphere and the ocean. In particular, it is closely linked to the surface energy budget due to its high albedo and its low thermal conductivity, which effectively cools the atmosphere. Additionally, sea ice caps the ocean temperature to the freezing point, and freezing salinizes whereas melting freshens the upper ocean, which alters the vertical density structure critical to water mass transformation. Air-sea ice interactions are not only thermodynamic; indeed, at first sight, pack ice seems motionless and rigid, when in reality it is moving horizontally, subject to winds and currents. Non-uniform drift generates major deformations ranging from large leads that expose the ocean directly to the atmosphere to pressure ridges that can be more than 20 m thick. The latter increase the mechanical coupling between the atmosphere and the ocean, increasing drag and generating turbulence in the air above and the water below. Depending on the conditions, pack ice can increase or decrease the transfer of momentum from the atmosphere to the ocean compared with an open ocean (Schulze and Pickart, 2012).

Pack ice is also a source of aerosols, whether airborne snow (Bergner et al., 2025), biogenic volatile compounds such as dimethyl sulphide or biogenic particles (bacterial or microalgal cell components), particularly in the presence of melt ponds where microorganisms sometimes abound and are in more direct contact with the atmosphere. When openings (leads) form in the pack ice, the ocean becomes a direct source of these aerosols, as well as moisture. These exchanges directly affect cloud cover through condensation or ice nuclei that, in turn, influence the surface energy budget and available radiation for photosynthesis.

The interface between pack ice and the ocean facilitates a variety of exchanges. First, the pack ice and its snow modulate the radiation that enters the ocean and is available for photosynthetic organisms, photoperception and underwater vision. As it grows, sea ice releases about 80% of the salts contained in freezing seawater to the ocean in the form of dense brines, generating convection and turbulent mixing in the water column. On a small scale near the ice-ocean interface, these motions help to replenish the internal channel network of the pack ice with nutrients essential for autotrophic organisms. The ice-ocean interface is also the site of a two-way exchange of microscopic and very small living organisms. A significant proportion of the organic carbon produced by these organisms in the pack ice is deposited in the water column, where it feeds certain marine organisms, is mineralised by the microbial cycle, or reaches the seafloor more or less intact.

*Targeted scales*

The scientific and deployment strategy adopted will enable us first of all to document vertical spatial scales, within, below and above the pack ice, and the seasonal temporal scale (Figure 1). We will be using new technologies to explore the interior of the sea ice on vertical scales from micrometres to centimeters to describe the microphysical structure relevant to microorganisms, as well as the stratigraphy of the sea ice. Above the pack ice, the distribution of aerosols and clouds over the first 2 km will be measured. Horizontally, measurements will take into account the variability in the properties of the pack ice from the metre to decametre scale, at the level of the snow cover and the ice, including those linked to deformation. On a larger scale, we will study variability linked to Atlantification and the long-range transport of air masses (100–1000+ km). Below the sea ice, at the top of the water column, we will be examining the scale of turbulent mixing due to shear and brine release, which affects the exchange of solutes and energy between the ice and the ocean. Turbulent mixing associated with deformations in the pack ice, such as compression ridge keels, will also be observed. The dynamics of the surface mixing layer (approximately 0–40 m) and the halocline (approximately 40–200 m) will be described, as will exchanges with the warm Atlantic Water layer below. The ocean (sub)mesoscale, e.g., in the form of eddies, will also be covered by the drifting platform. To capture the rapid transitions in seasonal variability, some sampling will be carried out on a daily or weekly basis. Around the equinoxes, the frequency will be increased to describe circadian variations. Finally, the sequence of multiple expeditions will be used to describe multiannual variability. In situ observations from the TPS will be combined with satellite observations to increase the spatial footprint. In addition, all observations will be integrated into a diversified modelling framework including specific processes and elements of the ecosystem (see Figures 2-6), the dynamics of the atmosphere, sea ice and the ocean, the radiation budget and, more broadly, the Earth System.

**THEMES**

A) *Life in the frozen ocean*

Arctic sea ice can be considered a major biome on Earth. Regardless of its age (hours to years) and state (slushy to hard and deep-freezing), it always harbours life. The porous nature of sea ice and the physical and chemical properties of its interstitial fluids control its habitability for a broad diversity of living organisms (Arrigo, 2017). Acting as a microorganism concentrator, sea ice provides confined micro-spaces where all ingredients for life from the atmosphere and the ocean combine (Babin et al., 2025). At times omnipresent across all latitudes during Earth's history (Hoffman et al., 1998), sea ice may well have been a hotspot for the evolution of microorganisms before life invaded the surface of continents. Today, it is a prominent component of polar marine ecosystems, providing a refuge for viruses, archaea, bacteria, protists, small invertebrates and fish larvae, despite its extreme and contrasting states (temperature, salinity, pH, alkalinity, gases) (Mundy and Meiners, 2021).

Many previous studies have addressed the phenology of the sea ice microbiome (SIM), i.e., the sequence of environmental changes and related biological responses over seasons (Lund-Hansen et al., 2020). Most, however, have focused solely on the spring bloom (Oziel et al., 2019), while only a few have addressed the polar night or the highly critical freeze-up period. In addition, they have most often focused on a single component of the SIM, whether it be microalgae, bacteria, viruses or protists, and on specific metabolic pathways. Furthermore, the role of micrograzers in the dynamics of the SIM has been largely neglected. From past studies, one may surmise that the SIM is a seasonally highly connected system in which what happens at any given time of the year under certain conditions will determine downstream events along the rest of the icy season (Lund-Hansen et al., 2020). For instance, colonisation of sea ice by microbes during freeze-up (Figure 3) and their survival strategies may be key in setting the stage for SIM dynamics during the spring bloom, with behaviour changeable from one year to another. Consequently, perhaps what happens at any given time in the SIM can only be fully understood in the context of an integrated and holistic seasonal conceptual framework, which should also account for the likely, but actually unknown, resilience of this complex system. At longer timescales, the tremendously increased physical proximity among microorganisms in sea ice may boost the rate of adaptation through horizontal gene transfer (Rapp et al., 2025) and provide additional resilience to the SIM. However, current knowledge of the SIM is too fragmentary for a solid holistic understanding of its dynamics and interactions which is necessary for predicting the fate of the Arctic SIM under current climate change, understanding the role that the SIM may have played during the evolution of life on Earth, and determining whether saline ice may harbour life in other ocean worlds.

Over a typical annual cycle, the sea ice ecosystem experiences dramatic transformations. The thermodynamic growth of sea ice largely drives the partitioning of gases, solutes and solids among variously shaped micro-spaces where microorganisms are found, creating a broad spectrum of living conditions, some being extreme and potentially representative of life on extraterrestrial ocean worlds. Our understanding remains patchy when it comes to how this ecosystem transitions over significant seasonal milestones, such as: i) **Freeze-up.** Sea ice is primarily colonised by microorganisms from the water column when the ice starts forming during autumn and thereafter as it grows. Strikingly, during the formation process, the density of living organisms is strongly enriched (a factor of up to 20,000) in sea ice compared with the underlying water column from which it derives (Gradinger and Ikävalko, 1998). What are the mechanisms responsible for sea ice colonisation by microorganisms (Figure 3)? Which microorganisms go on to survive the initial inclusion in sea ice? Which survival strategies allow sustained survival? How does the weather (temperature, wind, sea state) during freeze-up in any given year shape the sea ice ecosystem? ii) **Polar night.** During the polar night at very high latitudes, sea ice microorganisms confront annual minima in light and temperature (Figure 4). Photoautotrophs must endure darkness while bacteria and archaea, chemolithoautotrophs or mixotrophs may thrive. How do phototrophs survive and remain active in the absence of light? What adaptations are critical during that season? What is the possible role of the polar night in establishing the communities in subsequent seasons? iii) **Early wakeup.** Some photoautotrophic microalgae may start growing at surprisingly low irradiances in the Arctic Ocean (Hoppe et al., 2024). What are the minimum light requirements for photosynthesis? How important is early autotrophy and/or mixotrophy in the inoculation of a later bloom? iv) **Bloom.** As light availability increases, intense primary production locally develops in sea ice. What makes sea ice microenvironments so favourable? What are the mechanisms involved in nutrient pumping to the underlying water column? To what extent is the pre-conditioning (nutrient and organic carbon) by previous phases critical for the spring bloom? v) **Collapse of the sea-ice ecosystem and seeding of the upper ocean.** With the return of solar radiation, sea ice warms internally, brine inclusions interconnect, snow melts, and meltwater eventually flush down through the ice, favouring the appearance of melt ponds at the surface. Meanwhile, the ice bottom gets ablated, transferring many sympagic microorganisms to the water column (Figure 6). Is sea ice actually a refuge for microorganisms inoculating seawater? What are the adaptations that allow a cosmopolitan lifestyle (e.g., resting stages, tolerance of osmotic and other shocks)?

During Tara Polaris I, and follow-up drifts, we intend to address each of the above questions, and more broadly, to connect the dots among seasons from the freeze-up to break-up. Special focus will be put on documenting the scales that matter for biogeochemical cycles and microorganisms, from the small scale of brine inclusions to that of deformations in sea ice, irregularities in snow cover, occurrence and distribution of melt ponds, leads, and ridges.

### B) *Epi- and mesopelagic life in an ice-covered ocean*

A significant part of the CAO is perennially covered by sea ice which, combined with the extreme light regime at these high latitudes, makes the underlying ocean a unique marine ecosystem. In this permanent ice zone of the CAO, however, ice and water masses are constantly drifting, and not in the same direction. The underlying water masses come from the Pacific at the surface and from the Atlantic at depth under the halocline. During their transit, they are sometimes ice-free and sometimes ice-covered. The pack ice in the permanent ice zone of the CAO in the Eurasian Basin essentially contains first-year ice that has formed around the Laptev Sea, much of which will be expelled through Fram Strait. The CAO should therefore not be seen as a static ice-covered ocean. Its three compartments (including the atmosphere) are in perpetual motion relative to each other. It is this highly dynamic system that the TPS will be sampling, in a mode that is neither Lagrangian, nor Eulerian. It will likely spend most time above latitude 85°N.

The pack ice on the ocean does not constitute a solid layer preventing exchanges between the ocean and the atmosphere, quite the contrary. While while pack ice and its snow cover limit certain exchanges of gases and particles, it contributes to and sometimes even amplifies the transfer of wind kinetic energy from the atmosphere to the ocean, causing turbulent mixing and Ekman transport (Schulze and Pickart, 2012). During its growth in winter, sea ice releases dense brines that generate turbulent mixing in the water column by haline convection, causing the mixing layer to deepen and regenerating surface nutrient stocks for the following phytoplankton growth season (Peralta-Ferriz and Woodgate, 2015). In turn, the ocean directly affects the thermodynamics and dynamics of sea ice. The heat fluxes between the Atlantic waters of the halocline and the surface are of particular importance, as their changes linked to the so-called Atlantification of the Arctic Ocean are partly responsible for the reduction in sea ice. Sea ice is a refuge and a breeding ground for a multitude of organisms, which move between sea ice and the water column at different times of the year, depending on the state of the pack ice, conditions in the pelagic zone, and ontogenetic cycles. The two environments inoculate each other.

The CAO is considered an 'oligotrophic' ecosystem with lower annual primary production and biomass compared to the surrounding shelf area, where most previous Arctic studies have been conducted (Sakshaug, 2004). In the Eurasian basin, light is thought to be the primary factor that constrains the magnitude of annual primary production (Codispoti et al., 2013). While nutrient concentrations are generally low, they do not limit primary production but rather set a limit on the carrying capacity of biomass in this ecosystem. High grazing pressures by local micro- and mesozooplankton populations as well as expatriate populations that have been advected laterally from the more productive shelf and ridge areas into the central deep basins are also thought to keep algal biomass low, giving rise to a net heterotrophic system (Olli et al., 2007).

In the CAO, sympagic production makes up a significant proportion of total primary production (Gosselin et al., 1997; Ardyna et al., 2013), but it is thought to be contracting with the reduction in pack ice, to the benefit of the pelagic ecosystem and to the detriment of benthic organisms (Wassmann and Reigstad, 2011). A large part of the sympagic production of organic carbon ends up sedimenting in the water column in the form of detrital matter, stressed microorganisms or healthy microorganisms. Most of the sedimenting material is consumed by grazers and degraded by the microbial loop; some reaches the sea floor and fuels benthic communities. Many questions remain open: will the CAO remain a low biomass, net heterotopic ecosystem in the future or will we observe an increase in net production and accumulation with the retreat of sea ice, as has been already observed in some shelf areas? How will the relative contributions of sea ice algae and phytoplankton to the region's primary production change in the future? What would be the cascading effects on lipid and carbon transfer to higher trophic levels if dominance of ice algal production and phytoplankton were to shift? What are the key processes controlling carbon fluxes in the CAO?

Changes in biodiversity and community composition are also inevitable with the decrease of sea ice and growing signs of Atlantification effects in the CAO. Advancing knowledge of the pelagic ecosystem of the CAO is particularly desired in light of the 16-year international agreement, signed in 2021, to ban commercial fishing in the CAO until a better understanding of ecosystem properties and dynamics is gained. Because the exact community composition, seasonal and interannual dynamics of epi- and mesopelagic fish and zooplankton remain unknown, it is impossible at the moment to monitor the rate of borealization in the CAO. How the changing climate impacts higher trophic levels through bottom-up processes also remains largely unknown, particularly across seasons.

Polaris I will integrate novel genomics, imaging, optical and acoustic approaches, along with more traditional oceanographic measurements, to address knowledge gaps related to the biodiversity of pelagic communities in the CAO, from microbes to fish and marine mammals, and their interactions with sympagic communities. Activities will include studies on biological rhythms, seasonal behavior and trophic interactions of key species across several trophic levels as well as the processes that drive carbon and nitrogen fluxes in this ecosystem.

C) *Microbial and climate feedback processes in the Arctic atmosphere*

The Arctic atmosphere is a driver of climate warming through changes in clouds and the energy budget (Shupe et al., 2022). It is also undergoing profound changes in response to climate change, including potential effects of the changing airborne microbiome that are very poorly understood (Beck et al., 2024). Polaris I will explore the dynamics of airborne microorganisms —bacteria, archaea, microalgae, fungi, and viruses— over the Arctic Ocean, aiming to provide key insights into the feedback mechanisms linking microbial communities and climate processes.

Microbial feedbacks in the Arctic atmosphere include the emission, transport, and deposition of airborne microorganisms, originating from the Arctic sea ice and ocean. These microorganisms are locally entrained into the atmosphere through processes such as blowing snow or sea spray aerosolization. Once airborne, they can interact with atmospheric constituents, influencing cloud formation, precipitation, and radiative forcing.

One critical feedback loop involves the role of microbes as ice-nucleating particles (INPs) (Figure 5). Certain bacterial taxa, such as *Pseudomonas* and *Sphingomonas*, produce surface proteins that facilitate ice formation at relatively high temperatures. These biological INPs can initiate low-level cloud ice formation at relatively warm temperatures (e.g., –10°C), altering cloud properties and influencing the surface energy balance over the Arctic Ocean (Creamean et al., 2022). In the Arctic, where cloud radiative effects are particularly important for surface temperatures, microbial INPs might affect cloud longevity and albedo, contributing to either warming or cooling depending on cloud phase partitioning (fraction of liquid water and ice) and seasonal timing. Such changes may also impact photosynthetically available radiation arriving at the sea ice-ocean surface.

Moreover, particles can also be transported over long distances from terrestrial ecosystems after, for example, wind-driven resuspension of soils/dust, or vegetational shedding. Atmospheric long-range transport connects distant ecosystems to the Arctic. Air mass back-trajectory analyses have linked microbial community structure to long-range transport from marine and continental regions. This transport introduces microorganisms that may not be locally sourced, effectively "importing" microbial functions into the Arctic system. These imported microbes could influence local biogeochemistry after deposition, particularly if they become active upon settling onto snow, ice, or surface waters.

Microbial emissions are also responsive to climate change-induced changes. As Arctic sea ice retreats and permafrost thaws, more microbial habitats are exposed and become biologically active. Such activity can lead to increased microbial emissions into the atmosphere, reinforcing the feedback cycle. For instance, increased sea spray emission due to less sea ice may enhance the aerosolization of marine microbes, some of which may act as INPs (DeMott et al., 2016).

Marine, snow and sea ice microbes also produce volatile organic compounds that can undergo gas-to-particle conversion and participate in the formation of cloud condensation nuclei. There is a need to characterize such biological sources relative to other natural sources, such as mineral dust, and anthropogenic sources transported from lower latitudes (Schmale et al., 2021).

The above discussion makes evident that microbial diversity and atmospheric composition need to be studied as a function of meteorological conditions, such as temperature and wind speed, that influence boundary layer dynamics and large-scale circulation. These abiotic factors, in turn, are influenced by broader climatic trends, such as Arctic amplification and shifts in atmospheric circulation patterns. Thus, the microbial-climate feedback loop is inherently dynamic: local microbial emissions in the CAO affect climate-relevant processes in the lower atmosphere, and changing climate alters microbial community dynamics. The long-term observations on the TPS will aim to document these trends.

In summary, Arctic atmospheric microbial communities are both influencers and responders in the climate system. They contribute to aerosol and cloud processes that modulate regional climate, while simultaneously being shaped by the changing Arctic environment. These feedbacks highlight the need to integrate microbiology into climate models to better understand and predict Arctic climate trajectories.

D) *Contaminant sources and fate in the changing CAO*

Although seemingly remote, the CAO is highly interconnected with the rest of the globe through physical, atmospheric, and hydrological pathways that serve as conduits for the long-range transport of contaminants. This connectivity is of increasing concern as climate change and intensifying anthropogenic activities expose the Arctic to a growing burden of pollutants.

One of the main vectors of CAO contamination is the global oceanic system. The Atlantic and Pacific inflows transport dissolved and particulate substances into the Arctic basin, including industrial chemicals and trace metals. These pollutants can be remobilized from sediments or originate from lower latitudes, becoming integrated into Arctic water masses and biogeochemical cycles (Macdonald et al., 2005). Ocean currents also play a role in the northward migration of microplastics and nanoplastics, which have now been detected in Arctic surface waters, sea ice, and benthic sediments (Peeken et al., 2018; Bergmann et al.).

Freshwater inflow from the surrounding continents represents another major source of chemical inputs to the CAO. Although the Arctic Ocean holds only approximately 1% of the global ocean volume, it receives about 10% of global river discharge (Holmes et al., 2012). Rivers such as the Lena, Ob, Yenisei, and Mackenzie carry not only nutrients and organic carbon, but also a wide spectrum of contaminants, including pesticides, mercury, hydrocarbons, and industrial waste products. These riverine pollutants are derived from agriculture, mining, waste disposal, and atmospheric deposition over large watersheds. The seasonal pulse of river discharge, amplified by permafrost thaw and snowmelt under warming conditions, may lead to increased contaminant mobilization in coming decades (McGovern et al., 2022).

The atmosphere plays a dominant role in the long-range transport of airborne contaminants to the Arctic. Pollutants such as black carbon, sulfate aerosols, persistent organic pollutants, and mercury are delivered to the region through meridional air mass movements from mid-latitude industrial centers (AMAP, 2021). For example, elemental mercury ($Hg^o$) can be transported efficiently in the atmosphere and undergoes photochemical transformations in the Arctic spring, leading to atmospheric mercury depletion events and subsequent deposition to snow and sea ice (Dastoor et al., 2022). During summer melt, deposited mercury can be released into the ocean and enter marine food webs, posing risks to Arctic wildlife and indigenous communities reliant on subsistence harvesting.

Emerging pollutants, such as micro- and nanoplastics, have gained increasing attention as they are now ubiquitously present in the Arctic environment. Transported via atmospheric fallout, river discharge, and ocean currents, these plastic particles can act as carriers for other toxic compounds, including heavy metals and hydrophobic organic pollutants (Peeken et al., 2018). Their small size facilitates ingestion by marine organisms, raising concerns about trophic transfer and ecosystem-wide effects. The accumulation of plastics in Arctic sea ice suggests that the ice acts as a temporal sink and redistribution mechanism, releasing contaminants as it melts.

In addition to these long-range sources, local emissions are contributing increasingly to Arctic contamination. As sea ice retreats, maritime traffic—commercial, industrial, and tourism-related—is expanding rapidly, bringing emissions of $NO_x$, $SO_2$, particulate matter, and combustion-related byproducts (Law et al., 2017). Offshore oil and gas activities also release polycyclic aromatic hydrocarbons (PAHs) and other operational discharges. On land, localized sources such as diesel combustion, residential heating, and waste incineration contribute to wintertime air pollution, particularly in Arctic settlements. These emissions are often trapped by atmospheric inversions during cold months, exacerbating human exposure and regional deposition.

Climate change is amplifying several of these contaminant pathways. For instance, the increase in boreal wildfires, triggered by rising temperatures and increased lightning activity, releases large quantities of mercury, black carbon, and organic pollutants into the atmosphere (Pelletier et al., 2022; Hasan et al., 2025). These emissions can be transported into the Arctic, contributing to both direct contamination and indirect effects through ice-albedo feedbacks. Permafrost thaw is also expected to mobilize previously frozen contaminants—such as legacy pesticides and mercury—into aquatic systems, adding a new dimension to Arctic pollution (Schuster et al., 2018).

Despite growing evidence of contaminant presence in the CAO, significant knowledge gaps remain regarding their seasonal dynamics. The Arctic environment is characterized by extreme seasonal variations in temperature, light availability, sea ice cover, river discharge, and biological activity, all of which can influence the transport, transformation, and bioavailability of contaminants. For instance, mercury deposition events predominantly occur in spring, while summer melt releases accumulated contaminants from snow, ice, and permafrost. Similarly, freshwater discharge and primary productivity peaks in summer may enhance the mobilization and biological uptake of pollutants. Yet, most observational data are limited to summer months, and winter processes remain poorly documented due to logistical challenges. Understanding these seasonal patterns is essential to assess exposure risks for Arctic ecosystems and communities year-round. Research on TPS will give new data on year-round monitoring and modeling of contaminant fluxes and transformations to capture the full complexity of Arctic pollution under ongoing climate change.

The CAO is not an isolated or pristine environment but a highly connected and vulnerable region subject to diverse and growing contaminant pressures. Understanding the complex interplay of long-range and local sources, climate feedbacks, and ecological impacts is critical for anticipating future risks and for designing targeted monitoring and mitigation strategies.

*E) The TPS long-term observatory*

Long-term time series are essential to detect and attribute changes related to global climate change in the Arctic Ocean. As shown by Henson et al. (2010) and (Cael et al., 2023), time series spanning 20 to 40 years are required to reliably separate anthropogenic trends from natural variability in marine biogeochemical cycles and ecosystems.

The TPS observatory, to be deployed in the CAO over the next two decades, is designed to meet this challenge. It will focus on a carefully selected set of sentinels of global change (Nicolaus et al., 2025), chosen to characterize the physical state of the ocean-ice-atmosphere system (e.g., meteorology, snow/ice cover, hydrography), biogeochemical cycles (e.g., nutrients, carbon), and the structure and function of ecosystems (e.g., biodiversity, microbial functioning and carbon export). Data collection will follow internationally harmonized protocols to ensure robustness, interoperability, and long-term scientific value.

Observing mechanisms of climate change, as well as detecting long-term climate signals in the Arctic, is particularly challenging due to the superposition of anthropogenic trends with natural variability on interannual to decadal scales (e.g., ENSO, NAO, AO, PDO), which can mask or mimic persistent changes. At the atmospheric level, the rate of warming will be influenced strongly by sea ice loss and by the response of the atmospheric vertical structure and properties, in particular temperature, humidity and clouds. Also, shifts in climate modes, such as the North Atlantic Oscillation (NAO), have altered circulation patterns, with cascading effects on ocean currents and ecosystems; for example, coccolithophore blooms in the Barents Sea, initially interpreted as a warming-driven expansion, were later attributed to NAO-driven current variability (Neukermans et al., 2018). Sea ice extent also illustrates this complexity: after a sharp decline from 7.5 million km² to 4.5 million km² between 1979 and 2007, it has since stabilized around that lower value, likely reflecting a regime shift associated with the loss of older, thicker multiyear ice (Sumata et al., 2023; Johannessen and Olaussen, 2025; Stern, 2025). At the base of the food web, (Li et al., 2009) documented a shift from nanophytoplankton to picophytoplankton in the Beaufort Gyre due to freshwater-induced stratification, but with an extended dataset, (Li et al., 2013) revealed a more complex and non-monotonic pattern, underscoring the limitations of short-term observations for capturing ecosystem change.

These examples highlight the necessity of sustained, long-term observational efforts to accurately interpret environmental change in the Arctic. In addition to this temporal commitment, success will depend on a rigorous and adaptive implementation strategy that ensures the sustained relevance and scientific value of the monitored variables. The TPS observatory, through its strategic design and multi-decadal perspective, will provide the essential data needed to disentangle long-term climate signals from natural variability in one of the fastest-changing and most poorly observed regions of the global ocean.

**Expected results and impacts**

The Tara Polaris programme aims are:

1. To bring about a quantum leap in our knowledge of the most fundamental aspects of life, particularly microbial life, in frozen marine environments, whether in terms of adaptations to the extreme light and temperature conditions of this environment, biological innovations linked to the constraints imposed by the microphysics of sea ice, or chronobiological phenomena specific to large seasonal variations in light. This new knowledge will enable us to better understand these ecosystems in their current state, at different periods in the history of our planet throughout the evolution of life on Earth, and possibly elsewhere on other planets with a liquid ocean covered by saline ice.
2. To better understand the role of living organisms in the dynamics of the coupled ocean-ice-atmosphere system. The Tara Polaris programme is the first to place such importance on the role of biology in this coupled system and will devote many new observational and experimental approaches to gaining an in-depth understanding of the nature of biology-climate interactions in the polar environment.
3. To extract robust long-term trends directly linked to global change (climate and pollution) from multi-year variability, thanks to a series of expeditions covering at least 20 years. Key variables of the ecosystem and, more broadly, of the Earth system will be targeted.

The Tara Polaris programme will also be integrated into concerted Ocean observation initiatives at European and international levels (e.g., Arctic Monitoring and Assessment Programme), and will establish relationships with space agencies (CNES, ESA, NASA), the Copernicus service, operational oceanography agencies (e.g., Mercator Ocean), weather forecasting agencies (e.g., Météo France) and climate modelling agencies (e.g., European Centre for Medium-Range Weather Forecasts).

**Implementation of the Tara Polaris programme**

This special collection of vision papers (Geoffroy et al., 2025; Ghiglione et al., 2025; Nicolaus et al., 2025; Schmale et al., 2025; Vancoppenolle et al., 2025) highlights gaps in knowledge, emerging questions and approaches that will be taken during the inaugural Tara Polaris I expedition (2026–2028) to characterize and better understand the different compartments in the coupled sea ice-ocean-atmosphere system of the CAO, and the linkages between them. The small size of the platform that can only accommodate a small team of scientists will necessitate a strong coordination between the activities of the different themes, eventually leading to a more integrative and holistic approach compared to past studies that have been conducted by individual groups on large icebreakers. Research from the platform will be carried out in a similar manner to the International Space Station, in which a small team of highly trained 'Taranauts' will carry out all the protocols for the different measurements identified for each theme. Advances in communication will ensure good coordination with the teams onshore who will provide technical support to the team onboard.

To achieve an integrative outcome, the Tara Polaris endeavor is organized as a consortium of laboratories governed by an executive committee, an international science advisory board and a strategic board. Each of the five themes is coordinated by a team of PIs who worked closely with the executive board and consortium members to develop the vision papers for this special Elementa collection and the implementation of the scientific program. The Tara Polaris consortium adopts an 'open science' framework and FAIR (findable, accessible, interoperable, reusable) data principles. The consortium is open for consideration of proposals for new questions and novel tools that can advance the scientific knowledge of the CAO ecosystem, and these will be evaluated by the executive board in consultation with theme leaders. Proposers should be mindful of the existing framework for Tara Polaris I (as described in this collection of papers), the time and human constraints onboard, and the fact that the first drift will be followed by a series of additional drifts (Polaris II, III, …) that will open up opportunities for new ideas and contributions.

0aeolian dust and ecosystems. *Frontiers in Environmental Science* **13**. doi:10.3389/fenvs.2025.1536395.

Morris, AD, Wilson, SJ, Fryer, RJ, Thomas, PJ, Hudelson, K, Andreasen, B, Blévin, P, Bustamante, P, Chastel, O, Christensen, G, Dietz, R, Evans, M, Evenset, A, Ferguson, SH, Fort, J, Gamberg, M, Grémillet, D, Houde, M, Letcher, RJ, Loseto, L, Muir, D, Pinzone, M, Poste, A, Routti, H, Sonne, C, Stern, G, Rigét, FF. 2022. Temporal trends of mercury in Arctic biota: 10 more years of progress in Arctic monitoring. *Science of The Total Environment* **839**: 155803. doi:https://doi.org/10.1016/j.scitotenv.2022.155803.

Moulin, C, Linkowski, T, Petit, O, Valette, L, Hertau, M, Marie, L, Lebredonchel, V, Le Helleix, Y, Havet, W, Regnacq, P, Deux, J-M, Benamer, N, Seguin, V, Troublé, R. 2025. Tata Polar Station: Platform design. *Elementa: Science of the Anthropocene* (submission anticipated).

Mundy, CJ, Meiners, KM. 2021. Ecology of Arctic Sea Ice, in Thomas DN ed., *Arctic Ecology*. John Wiley & Sons: 261-288.

Nansen, F. 1902. *The Oceanography of the North Polar Basin*. Longmans, Green, and Company.

Neukermans, G, Oziel, L, Babin, M. 2018. Increased intrusion of warming Atlantic water leads to rapid expansion of temperate phytoplankton in the Arctic. *Global Change Biology* **24**(6): 2545-2553. doi:10.1111/gcb.14075.

Nicolaus, M, Ardyna, M, Houssais, M-N, Raut, J-C, Bisson, K, Flores, JM, Galand, PE, Geoffroy, M, Heimbürger-Boavida, L-E, Law, KS, Lovejoy, C, Ravetta, F, Rysgaard, S, Schmale, J, Schuback, N, Sonke, JE, Vancoppenolle, M, Tremblay, J-É, Babin, M, Bowler, C, Karp-Boss, L, Troublé, R. 2025. Tara Polar Station: Sustained decadal observations of the coupled Arctic system in rapid transition. *Elementa: Science of the Anthropocene* (submitted).

Nicolaus, M, Gerland, S, Hudson, SR, Hanson, S, Haapala, J, Perovich, DK. 2010. Seasonality of spectral albedo and transmittance as observed in the Arctic Transpolar Drift in 2007. *Journal of Geophysical Research: Oceans* **115**(C11). doi:https://doi.org/10.1029/2009JC006074.

Nicolaus, M, Perovich, DK, Spreen, G, Granskog, MA, von Albedyll, L, Angelopoulos, M, Anhaus, P, Arndt, S, Belter, HJ, Bessonov, V, Birnbaum, G, Brauchle, J, Calmer, R, Cardellach, E, Cheng, B, Clemens-Sewall, D, Dadic, R, Damm, E, de Boer, G, Demir, O, Dethloff, K, Divine, DV, Fong, AA, Fons, S, Frey, MM, Fuchs, N, Gabarró, C, Gerland, S, Goessling, HF, Gradinger, R, Haapala, J, Haas, C, Hamilton, J, Hannula, H-R, Hendricks, S, Herber, A, Heuzé, C, Hoppmann, M, Høyland, KV, Huntemann, M, Hutchings, JK, Hwang, B, Itkin, P, Jacobi, H-W, Jaggi, M, Jutila, A, Kaleschke, L, Katlein, C, Kolabutin, N, Krampe, D, Kristensen, SS, Krumpen, T, Kurtz, N, Lampert, A, Lange, BA, Lei, R, Light, B, Linhardt, F, Liston, GE, Loose, B, Macfarlane, AR, Mahmud, M, Matero, IO, Maus, S, Morgenstern, A, Naderpour, R, Nandan, V, Niubom, A, Oggier, M, Oppelt, N, Pätzold, F, Perron, C, Petrovsky, T, Pirazzini, R, Polashenski, C, Rabe, B, Raphael, IA, Regnery, J, Rex, M, Ricker, R, Riemann-Campe, K, Rinke, A, Rohde, J, Salganik, E, Scharien, RK, Schiller, M, Schneebeli, M, Semmling, M, Shimanchuk, E, Shupe, MD, Smith, MM, Smolyanitsky, V, Sokolov,
29aeolian dust and ecosystems. *Frontiers in Environmental Science* **13**. doi:10.3389/fenvs.2025.1536395.

Morris, AD, Wilson, SJ, Fryer, RJ, Thomas, PJ, Hudelson, K, Andreasen, B, Blévin, P, Bustamante, P, Chastel, O, Christensen, G, Dietz, R, Evans, M, Evenset, A, Ferguson, SH, Fort, J, Gamberg, M, Grémillet, D, Houde, M, Letcher, RJ, Loseto, L, Muir, D, Pinzone, M, Poste, A, Routti, H, Sonne, C, Stern, G, Rigét, FF. 2022. Temporal trends of mercury in Arctic biota: 10 more years of progress in Arctic monitoring. *Science of The Total Environment* **839**: 155803. doi:https://doi.org/10.1016/j.scitotenv.2022.155803.

Moulin, C, Linkowski, T, Petit, O, Valette, L, Hertau, M, Marie, L, Lebredonchel, V, Le Helleix, Y, Havet, W, Regnacq, P, Deux, J-M, Benamer, N, Seguin, V, Troublé, R. 2025. Tata Polar Station: Platform design. *Elementa: Science of the Anthropocene* (submission anticipated).

Mundy, CJ, Meiners, KM. 2021. Ecology of Arctic Sea Ice, in Thomas DN ed., *Arctic Ecology*. John Wiley & Sons: 261-288.

Nansen, F. 1902. *The Oceanography of the North Polar Basin*. Longmans, Green, and Company.

Neukermans, G, Oziel, L, Babin, M. 2018. Increased intrusion of warming Atlantic water leads to rapid expansion of temperate phytoplankton in the Arctic. *Global Change Biology* **24**(6): 2545-2553. doi:10.1111/gcb.14075.

Nicolaus, M, Ardyna, M, Houssais, M-N, Raut, J-C, Bisson, K, Flores, JM, Galand, PE, Geoffroy, M, Heimbürger-Boavida, L-E, Law, KS, Lovejoy, C, Ravetta, F, Rysgaard, S, Schmale, J, Schuback, N, Sonke, JE, Vancoppenolle, M, Tremblay, J-É, Babin, M, Bowler, C, Karp-Boss, L, Troublé, R. 2025. Tara Polar Station: Sustained decadal observations of the coupled Arctic system in rapid transition. *Elementa: Science of the Anthropocene* (submitted).

Nicolaus, M, Gerland, S, Hudson, SR, Hanson, S, Haapala, J, Perovich, DK. 2010. Seasonality of spectral albedo and transmittance as observed in the Arctic Transpolar Drift in 2007. *Journal of Geophysical Research: Oceans* **115**(C11). doi:https://doi.org/10.1029/2009JC006074.

Nicolaus, M, Perovich, DK, Spreen, G, Granskog, MA, von Albedyll, L, Angelopoulos, M, Anhaus, P, Arndt, S, Belter, HJ, Bessonov, V, Birnbaum, G, Brauchle, J, Calmer, R, Cardellach, E, Cheng, B, Clemens-Sewall, D, Dadic, R, Damm, E, de Boer, G, Demir, O, Dethloff, K, Divine, DV, Fong, AA, Fons, S, Frey, MM, Fuchs, N, Gabarró, C, Gerland, S, Goessling, HF, Gradinger, R, Haapala, J, Haas, C, Hamilton, J, Hannula, H-R, Hendricks, S, Herber, A, Heuzé, C, Hoppmann, M, Høyland, KV, Huntemann, M, Hutchings, JK, Hwang, B, Itkin, P, Jacobi, H-W, Jaggi, M, Jutila, A, Kaleschke, L, Katlein, C, Kolabutin, N, Krampe, D, Kristensen, SS, Krumpen, T, Kurtz, N, Lampert, A, Lange, BA, Lei, R, Light, B, Linhardt, F, Liston, GE, Loose, B, Macfarlane, AR, Mahmud, M, Matero, IO, Maus, S, Morgenstern, A, Naderpour, R, Nandan, V, Niubom, A, Oggier, M, Oppelt, N, Pätzold, F, Perron, C, Petrovsky, T, Pirazzini, R, Polashenski, C, Rabe, B, Raphael, IA, Regnery, J, Rex, M, Ricker, R, Riemann-Campe, K, Rinke, A, Rohde, J, Salganik, E, Scharien, RK, Schiller, M, Schneebeli, M, Semmling, M, Shimanchuk, E, Shupe, MD, Smith, MM, Smolyanitsky, V, Sokolov,

**Acknowledgements**

MB was supported by Sentinel North, Québec-Océan and FRQNT. This article is a contribution to the Norwegian Research Council project 352539 (SEDNA). JMF is supported by the generous contributions of Scott Eric Jordan. SR was supported by Aage V Jensens Foundations (grant# 2021-12-30) and the Danish National Research Foundation (grant# DNRF 185). J.S. was supported by the Swiss National Science Foundation (grant# 200021_212101) and holds the Ingvar Kamprad Chair for extreme environments research, sponsored by Ferring Pharmaceuticals.

**Author contributions**

Substantial contributions to conception and design
All

Drafting the article or revising it critically for important intellectual content
All

Final approval of the version to be published
　　All

**List of figures**

**Figure 1.** **Schematic of spatial/temporal scales and key physical features to be captured during the Tara Polaris expeditions.**

Horizontally, time of the year over more than a year is depicted in terms of solar-noon sun elevation, lightscape (shades of blue in the atmosphere), and equinoxes and solstices dates (yellow rectangles). Seasonal changes in clouds, the state of the pack ice, and the physical properties of the ocean are also depicted. In the ocean, the three horizontal dashed lines show the approximate depth (see vertical axis) of the surface mixed layer, the cold halocline and the lower halocline. Seasonal variations are captured for the surface mixed layer. The circular arrows represent the turbulent mixing associated with the keel of pressure ridges, which are moving relative to the underlying water masses. Surface currents due to sea ice drift stress and large-scale currents are also shown. Although not shown, sub-mesoscale phenomena such as eddies are likely to be captured, as the drifting platform will provide quasi-synoptical sampling of such physical features. Seasonal changes in the pack ice include, from left to the right, the open-water period, the freeze-up, thermodynamic growth of sea ice to maximum thickness in spring, deformations due to dynamics (pressure ridges and leads), the melt season with appearance of melt ponds, and the ice break-up. The snow cover appears early after freeze-up and disappears during the melt season. In the atmosphere, low- and high-level clouds are shown, with their respective altitude (see vertical axis), and a representation of seasonal variations in their occurrence and height. The wind blowing at the surface and promoting aerosolization, especially above open water, including open leads, is represented. The three horizontal axes depict the small, medium and large scales that will be captured by the TPS sampling strategy described in detail in Geoffroy et al. (2025), Ghiglione et al. (2025), Nicolaus et al. (2025), Schmale et al. (2025) and Vancoppenolle et al. (2025). Small scale features include variations in sea ice and snow cover thicknesses, pressure ridges, turbulent mixing, and melt ponds. Medium-scale features include sea ice concentration, leads, eddies, and aerosols under heterogeneous sea ice conditions (marginal ice zone and leads). Large scale features include changes in water and air masses, and long-range transport of aerosols, contaminants and pollutants. Variations in biological and biogeochemical properties associated with all these features will be sampled.

**Figure 2.** **Schematic of zooplankton distribution with depth and time of year in the central Arctic Ocean.**

Zooplankton is represented by *Calanus* sp. and *Othonia* sp. Arctic cod is presented at various life stages (larva, juvenile and adult). The vertical double-head arrows in the ocean represent the amplitude of the diel vertical migrations as a function of the time of year. Please refer to Figure 1 for details of other features. The four rectangles (A, B, C and D) point to specific ecosystems and time of year, with further details provided in Figures 3–6 about some of the phenomena that will be the subject of study.

**Figure 3.** **Illustration of four potential colonization mechanisms by which microbes enter sea ice.**

Scavenging takes place while frazil ice is present and vertically stirred. Bacteria and microalgae may use specific molecules such as ice-binding proteins to adhere to frazil grains and thereby become trapped in ice when frazil ice coalesces and consolidates to form granular ice. Ice nucleation may be triggered by microbes in contact with supercooled water at the ocean surface during the freeze-up. Ice-binding proteins may be used by the microorganisms to trigger congelation of supercooled water at even warmer temperatures. Motility, thanks to the use of flagella or buoyancy adjustment, may be other ways for microbes to enter or be entrained into sea ice. See (Babin et al., 2025) for a more detailed description of these and other colonization mechanisms.

**Figure 4.** **Illustration depicting the sea ice ecosystem at the microbial scale.**

The brine channels within the ice matrix are depicted, as are the extreme conditions under which the microbes live (low temperature, low light and high salinity). Viruses, bacteria, archaea (not graphically shown but implicitly present) and microalgae are also shown. The lighter shade of blue at the ice-water interface represents the presence of antifreeze molecules produced by microbes, such as exopolymeric substances (EPS).

**Figure 5.** **Illustration showing interactions between biological aerosols and low-level clouds in the central Arctic Ocean.**

Biological aerosols (living or dead microbes or microbial debris) may originate locally from sea spray or melt ponds, or be transported to the Arctic in air masses originating from lower latitudes. The way in which they are distributed vertically in the atmosphere, affects their interaction with supercooled water and ice crystals contained in clouds. Abiotic particles were excluded for graphical simplicity. Here, biological aerosols can act as ice-nucleating particles, particularly thanks to the production of ice-binding proteins, or as cloud condensation nuclei. The 1-km vertical scale refers to the atmospheric boundary layer.

Figure 6. **Illustration showing different fates for organic matter released from pack ice during the spring melt.**

Three pathways are shown: 1) inoculation of living microorganisms into the water column ("seeding" the plankton), 2) grazing of released microbes and detritus by zooplankton at different depths, and 3) sedimentation to the lower water layers and seafloor. During transit to the bottom, organic matter also undergoes degradation by bacterial activity.

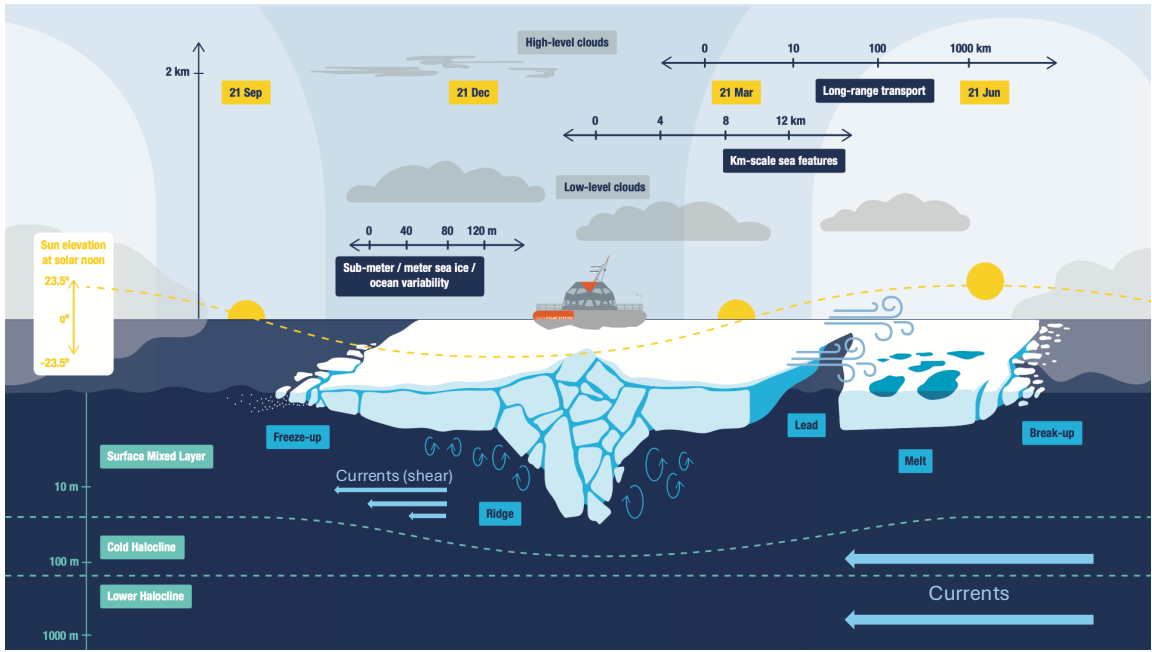

Figure 1

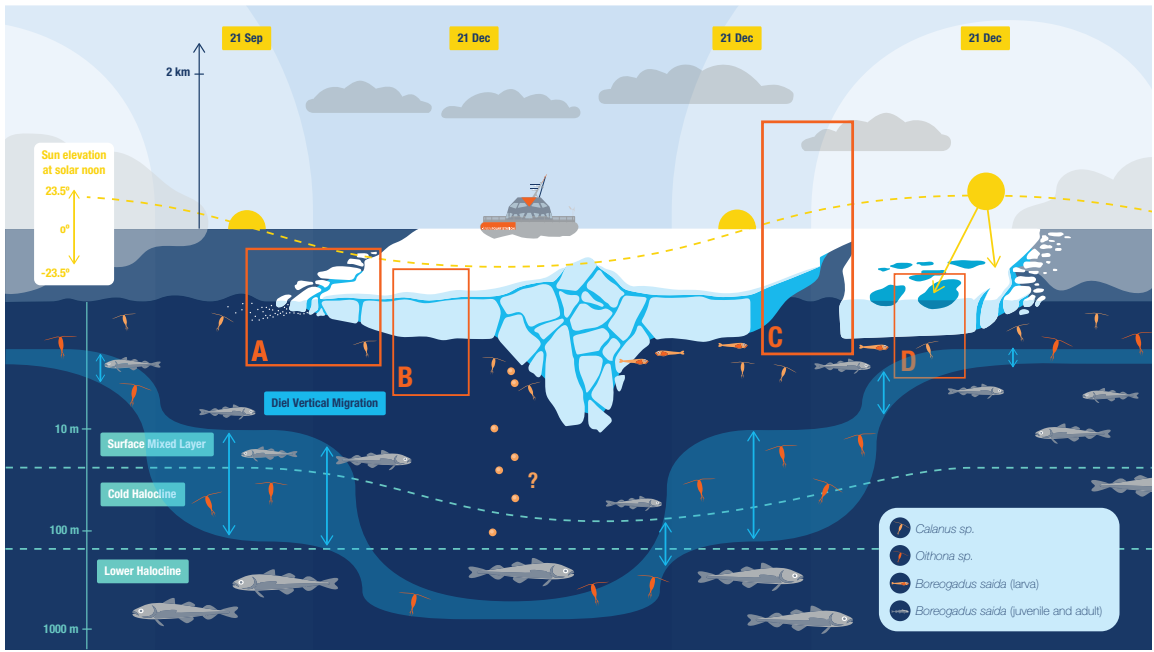

Figure 2

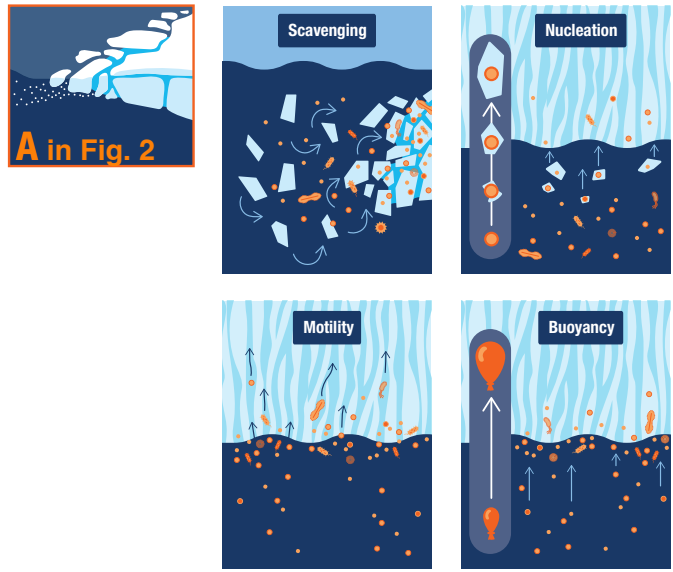

Figure 3

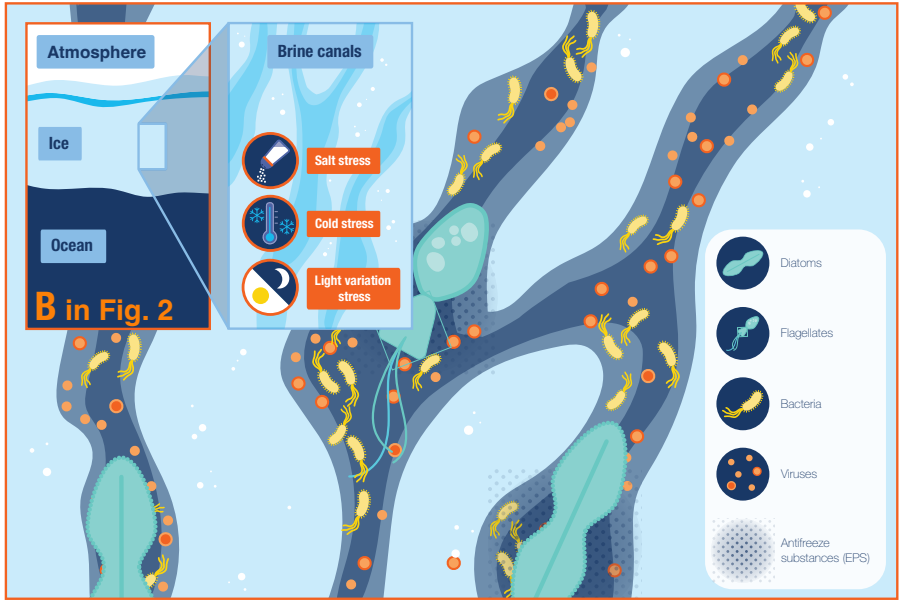

Figure 4

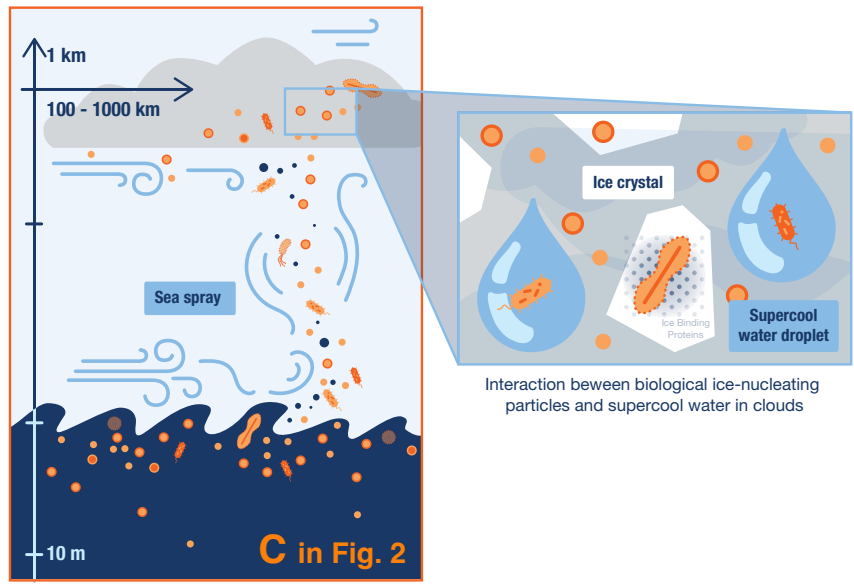

Figure 5

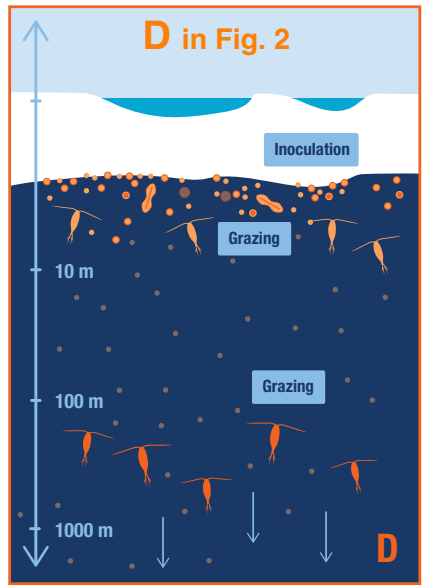

Figure 6